%\mag=\magstephalf
\mag=\magstep1
\pageno=1
\input amstex
\documentstyle{amsppt}
\TagsOnRight
\NoRunningHeads

\pagewidth{16.5 truecm}
\pageheight{23.0 truecm}
\vcorrection{-1.0cm}
\hcorrection{-0.5cm}
%\advance\vsize by -\voffset
%\advance\hsize by -\voffset
%\baselineskip = 0.8 true cm
\nologo

\NoBlackBoxes
\font\twobf=cmbx12

\define \ee{\roman e}

\define \RR{{\Bbb R}}

\define \ZZ{{\Bbb Z}}

\define \BK{Berry \& Klein 1996}
\define \BKT{1996}
\define \BB{Bohr 1928}
\define \BW{Born \& Wolf 1975}
\define \BU{Berry \&  Ustill 1980}
\define \GS{Guillemin \& Sterngerg 1984}
\define \HB{Hannay \& Berry 1980}
\define \HBT{1980}
\define \Kl{Klein 1926}
\define \He{Hecke 1981}
\define \IMO{Ishiwata, Matsutani \& \^Onishi 1997}
\define \IR{Ireland \& Rosen 1990}

\define \Mas{Maslov  1972}
\define \MatO{Matsutani 1997}
\define \MatTT{Matsutani 2001a, 2001b}

\define \NT{Noponen \& Turunen 1993}
\define \NTT{1993}
\define \Pat{Patorski 1989}
\define \Sch{Schulman 1981}
\define \SW{S\'anchez \& Wolf 1985}
\define \Tan{Tanaka 1986}

\define \TalT{1838}
\define \W{Weil 1964}
\define \WW{Winthrop \& Worthington 1965}
\define \WWT{1965}

%\define \tvskip{\vskip 0.5 cm}
\define \tvskip{\vskip 1.0 cm}
\define\ce#1{\lceil#1\rceil}
\define\dg#1{(d^{\circ}\geq#1)}
\define\Dg#1#2{(d^{\circ}(#1)\geq#2)}

\define\s#1{\sigma_{#1}}
\define\tp#1{\negthinspace\left.\ ^t#1\right.}
\define\mrm#1{\text{\rm#1}}
\define\lr#1{^{\sssize\left(#1\right)}}

{\centerline{\bf{Wave-Particle Complementarity
 and Reciprocity of Gauss Sums}}}

{\centerline{\bf{on Talbot Effects}}}

\author
Shigeki MATSUTANI${}^*$ and
 Yoshihiro \^ONISHI${}^\dagger$
\endauthor
\affil
${}^*$8-21-1 Higashi-Linkan Sagamihara, 228-0811 Japan \\
${}^\dagger$Faculty of Humanities and Social Sciences,
 Iwate University,\\
  Ueda, Morioka, Iwate, 020-8550 Japan
\endaffil \endtopmatter

\centerline{\twobf Summary }
\vskip 0.5 cm

Berry and Klein (J. Mod. Opt. (1997) 43 2139-2164)
showed that the Talbot effects in classical optics
are naturally expressed by Gauss sums in number theory.
Their result was obtained by a computation of Helmholtz equation.
In this article, we calculate the effects using Fresnel integral
and show that the result is also represented by Gauss sums.
However function forms of these two computational results 
are apparently different. 
We show that  the reciprocity law of Gauss sums
connects  these results and both completely agree with.
The Helmholtz equation can be regarded as an equation
based upon wavy nature in optics whereas
the Fresnel integral is defined by a sum over
the paths based upon a particle picture in optics.
Thus the agreement of these two computational results
could be interpreted in terms of the concept of 
the wave-particle complementarity,
though the concept is for quantum mechanical phenomenon.
This interpretation leads us to a 
relation between the reciprocity of Gauss sums in number theory
and  the wave-particle complementarity in wave physics.
We  discuss it in detail.

%\subheading{PACS numbers}:
\tvskip

\centerline{\twobf Keywords }

Talbot Effects, Gauss Sums, Complementarity, Reciprocity

%\newpage
\document
%\baselineskip= 0.8 true cm

\vskip 1 cm
{\centerline{\bf{\S 1. Introduction}}}
\vskip 0.5 cm

The Talbot effects are  diffraction grating
phenomenon in classical optics, discovered by Talbot (\TalT),
which are known as self-image (\Pat);
due to the effects, same
patterns arrayed on one or two dimensional grating
sharply recover on the screen
for a certain condition without lens system.
It is remarkable that
such the effects occurs due to wavy properties.

Recently Berry and Klein (\BKT) discovered that behind
the phenomenon there is an arithmetic structure.
They considered optical wave
 with wavelength $\lambda$
which stems from a one-dimensional grating with a period $a$.
The grating is set up so that its plane agrees with
the wave front and the screen is set parallel
to the grating with distance $z$.
They investigated Talbot effects
when $z$ was a fractional number times the Talbot distance
$z_T:=a^2/\lambda$.
They found a beautiful connection between wave physics
and number theory in the investigation of the Talbot effects,
which is based upon a deep insight by Hannay and Berry (\HBT).
Berry and Klein investigated
the Talbot effects by solving the Helmholtz equation (\NT)
and showed that the Talbot effects
were explained by Gauss sums and the
quadratic reciprocity law (see for example \IR, Chap. 6).

In this article, we present
another connection of the problem with Gauss sums
based upon the arguments of Berry and Klein.
We will also neglect the polarization effect
(\NT) and partially review  investigation of Berry and Klein
by solving {\it the  Helmholtz equation} in \S 2.
In \S 3, we compute the same system by means of
{\it Fresnel integral} (\WW), whose apparent function form
differs from the results in \S 2.
However  in terms of the reciprocity of Gauss sums
we will show the agreement
of both results in \S 4. We should note that
whereas Helmholtz equation is a wave equation,
the Fresnel integral is an integration over the optical paths
with a natural integration weight. The former one should be regarded as
an expression of the wavy properties and later one should be
interpreted as a transformation to wave expression
from the particle nature.
Thus the agreement of the both results
reminds us of wave-particle complementarity, {\it i.e.},
the wave-particle duality of nature in the
quantum mechanics (\BB).
We also argue
 that the complementarity is related to
the reciprocity of Gauss sums (\He\  Chap.8) in the Talbot effects,
by using the analogy between
the quantum mechanics and the optics in \S 4.

In order to make the agreements between two computations
from Helmholtz equation and Fresnel integral
easier,
we sometimes employ different expressions from those
in the papers of Berry and Klein and Hannay and Berry
({\it loc.cit.}).
Thus the difference between our formula and theirs
sometimes occurs in this paper.

Interestingly in the same period as Talbot's discovery,
Gauss lived and studied optics,
number theory including
Gauss sums and quadratic residues and so on (\Kl).

\vskip 1 cm
{\centerline{\bf{\S 2. Wavy Expressions}}}
\vskip 0.5 cm

As in the paper of Berry and Klein (\BKT),
we will start with an incident plane wave
of wavelength $\lambda$  coming
through periodic
$\delta$ functions transparency whose period is $a$.
Such a grating is called $\delta$-comb.
Let the transverse direction denote
$x$ while $z$ denotes along the optical axis.

In this section, we will deal with the system in terms of
the Helmholtz {\it wave} equation following
Berry and Klein (\BKT) and
Noponen and Turunen(\NTT).
Let us introduce the dimensionless parameters with respect to
Talbot distance $z_T$,
$$
        \xi := \frac{x}{a}, \quad
        \zeta := \frac{z}{z_T} = \frac{z\lambda}{a^2}.
\tag 2-1
$$
On the $\delta$-comb grating plane $\zeta=0$, the optical wave
 is expressed by
the Poisson sum formula,
$$
        \psi_{comb}(\xi,0)=\sum_{n = -\infty}^\infty \exp(2\pi i\xi n)
                    =\sum_{m = -\infty}^\infty \delta(\xi-m).
 \tag 2-2
$$
As there is a discrete translation symmetry, we can set $\xi \in \RR$
as $\xi = \xi_0 + n $ for $\xi_0 \in (-1/2, 1/2)$ and an integer $n$.
By letting
$$
        \psi_{comb}(\xi,\zeta)=\sum_{n = -\infty}^\infty
                 \eta(\zeta)\exp(2\pi i\xi n),
    \tag 2-3
$$
we substitute it into the Helmholtz equation,
$$
        \left( \frac{\partial^2}{\partial x^2}
              + \frac{\partial^2}{\partial z^2}
              + \left(\frac{2\pi}{\lambda} \right)^2\right)
 \psi_{comb} = 0.
      \tag 2-4
$$
Then we obtain the solution of (2-4) up to a constant
factor $C$,
$$
        \psi_{comb}(\xi,\zeta)=C\sum_{n = -\infty}^\infty
         \exp(2\pi i\xi n)
      \exp\left(2\pi i\zeta \left(\frac{a}{\lambda}\right)^2
    \left[1- \left(\frac{n\lambda}{a}\right)^2
       \right]^{1/2}\right) . \tag 2-5
$$
Here we note $\delta(cx) = \delta(x)/c$. For the case $n>a/\lambda$,
the exponential factor becomes a damping function.

By employing the paraxial approximation,
(2-5) is expressed by,
$$
        \psi_{comb}(\xi,\zeta)\approx C\sum_{n = -\infty}^\infty
         \exp(2\pi i\xi n)
           \exp\left(2\pi i\zeta \left(\frac{a}{\lambda}\right)^2
        \left[1- \frac{1}{2}\left(\frac{n\lambda}{a}\right)^2
          \right]\right). \tag 2-6
$$
Of course, large $n$ is out of the approximation but
as it is expected that the contribution from the large $n$
is less than that from small $n$ in Fourier analysis, we will go on
to consider (2-6) as in the paper of Berry and Klein (\BKT).
Let  the right hand side of (2-6) be denoted by $\psi_p(\xi,\zeta) \exp(i k z)
$
 following the notations in the paper, where $k := 2\pi/\lambda$.
We obtain,
$$
        \psi_p(\xi,\zeta)=C\sum_{n = -\infty}^\infty
                 \exp(2\pi i\xi n)
                 \exp\left(-2\pi i\zeta \left( \frac{1}{2}n^2\right)
                \right) . \tag 2-7
$$

Let us consider the case that $\zeta$ is a rational
number, {\it i.e.}, for coprime positive numbers $p$ and $q$
$$
        \zeta = \frac{p}{q}. \tag 2-8
$$
By introducing the integers $l$, $s$ such that $n=lq+s$, the
quantity of each term in (2-7) become,
$$
        \xi n - \frac{1}{2}n^2 \zeta= \xi s -
                \frac{1}{2}\frac{p}{q}s^2
              + \xi l q - \frac{1}{2} l^2 q p - lps.
         \tag 2-9
$$
Further we note the relation,
$$
        \ee^{-\pi i q p  l^2} = (-1)^{ql^2p}= \ee^{-\pi i q  p l }
      , \tag 2-10
$$
and it is the unity if $pq$ is even.
Then we have
$$
        \psi_p(\xi,\zeta)=C \sum_{s = 0}^{q-1}
                 \ee^{\pi i\xi (2 \xi q s - p s^2)/q }
           \sum_{l = -\infty}^\infty
                 \ee^{\pi i (2 \xi q  - p q) l },
         \tag 2-11
$$
Using the Poisson sum formula again, we obtain,
$$
        \psi_p(\xi,\zeta)=C\sqrt{q}\sum_{n= -\infty}^\infty
             A(n; q, p ) \delta (\xi -\frac{1}{2}e_{qp}
           - \frac{n}{q}  ),
          \tag 2-12
$$
where
$$
        e_{qp} := \left\{\matrix 1, &\text{if } qp & \text{odd,} \\
                0, &\text{if } qp & \text{even,} \endmatrix \right.
           \tag 2-13
$$
$$
         A(n; q, p )=\frac{1}{\sqrt{q}}
              \sum_{s = 0}^{q-1}
                 \exp\left({i\pi
         \left[\left( 2n+q e_{qp}\right)
             s - p s^2\right]/q }\right).
           \tag 2-14
$$

We note that if $pq$ is odd, $( 2n+q e_{qp})$ is also odd.
Hence if we can use the formula given by (A-18) in the Appendix
which is based on the Appendix of the paper of Hannay and Berry (\HBT),
then (2-14) is expressed as follows (\BK);
$$
A(n;q,p) =
\left\{ \matrix
\pmatrix p\\ q\endpmatrix
\exp\left(i
\pi \left[ \dfrac{1}{4}(q-1)
         + \dfrac{p}{q}
             \left(\left[\dfrac{1}{p}\right]_{q}\right)^2n^2
  \right]\right), &
             \text{  $p$ even, $q$ odd, } \\
\pmatrix q\\ p\endpmatrix\exp\left(- i
\pi \left[\dfrac{1}{4}p-
 \dfrac{p}{q}
  \left(\left[\dfrac{1}{p}\right]_{q}\right)^2 n^2
               \right] \right), &
             \text{  $p$ odd, $q$ even, } \\
\pmatrix p\\ q\endpmatrix \exp\left(i
\pi \left[ \dfrac{1}{4}(q-1)
         + \dfrac{2p}{q}
\left[\dfrac{1}{2}\right]_{q}
  \left(\left[\dfrac{1}{2p}\right]_{q}\right)^2
      (2n+q)^2
             \right]\right), &
             \text{  $p$ odd, $q$ odd.}\\
\endmatrix
\right.\tag 2-15
$$
where $\left[\dfrac{1}{p}\right]_{q}$ is a unique positive
integer smaller than $q$ satisfying
$$
p \left[\dfrac{1}{p}\right]_{q}\equiv 1
\text{  mod }q, \tag 2-16
$$
and $\pmatrix p \\ q\endpmatrix $ is the Jacobi symbol
which is a product of
the  Legendre symbol $ \pmatrix p \\ s \endpmatrix$
for the prime factors $s$ of $q$ (\IR, Chap. 5),
$$
\pmatrix p \\ s \endpmatrix :=
\left\{ \matrix +1, & \text{
if there is an integer $m$ such that $m^2 = p$ mod $s$,}\\
-1, & \text{otherwise.} \endmatrix \right.
\tag 2-17
$$

\vskip 1 cm
{\centerline{\bf{\S 3. Particle Expressions}}}
\vskip 0.5 cm

In this section, we will consider the system
in terms of the Fresnel integration
following the study by Winthrop and Worthington (\WWT).
Here we note that
the Fresnel integration should be considered as  a simple
path integration (\Sch, Chap.20; \SW) over the shortest optical paths
obeying the minimal principle of Fermat
and these paths should be regarded as a particle
picture in optics.

Using the Fresnel integral for $\delta$-comb wave function
(2-2), we obtain the $\psi_{comb}$
at the point $(\xi, \zeta)$,
$$
\tilde\psi_{comb}(\xi,\zeta) = \int
 \frac{a d \xi'}{\sqrt{ (\xi-\xi')^2a^2 + z^2 }}
\exp\left(      \frac{2\pi i}{\lambda}\sqrt{ (\xi-\xi')^2a^2 + z^2 }
\right) \psi_{comb}(\xi',0)\left(\cos\theta +1 \right),
\tag 3-1
$$
where $\cos\theta = z/\sqrt{ (\xi-\xi')^2a^2 + z^2 }$.
Here we note that (3-1) is a solution of (2-4) in the sense of
 $\Cal O(\lambda/ \sqrt{ (\xi-\xi')^2a^2 + z^2 })$ as
the function is an approximate kernel of the Helmholtz
differential operator, called Fresnel approximation
 \cite{BW, Chap.8}.

The optical distance between a point $(\xi , \zeta)$
and the $n$-th split $(-n, 0)$ is given by
$$
        \frac{1}{\lambda}\sqrt{ (\xi + n)^2a^2 + z^2 }, \tag 3-2
$$
and in the paraxial approximation, it is approximated by
$$
        \frac{1}{\lambda}
         z + \frac{1}{2} n^2 \frac{1}{\zeta} + \frac{\xi}{\zeta}n
           +\frac{1}{2} \frac{\xi^2}{\zeta}.
           \tag 3-3
$$
In this approximation,
the denominator of the first factor
is approximated by $z$ and $\cos\theta$ can be
regarded as the unity,
(3-1) becomes
$$
        \tilde\psi_{comb}(\xi,\zeta)\approx C'
          \sum_{n = -\infty}^\infty
                 \exp(2\pi i z /\lambda)
                 \exp\left(\pi i\left(\frac{2\xi n}{\zeta}
                 + \frac{n^2}{\zeta}+
               \frac{\xi^2}{\zeta}\right)\right),
           \tag 3-4
$$
using a certain constant factor $C'$.
Of course, this approximation is not contradict with the Fresnel approximation
.
By letting the right hand side of (3-4)
denoted by $\tilde\psi_p\exp(i k z)$,
we have
$$
        \tilde\psi_p(\xi,\zeta)=C'
              \sum_{n = -\infty}^\infty
           \exp\left(\pi i\left(\frac{2\xi n}{\zeta}
                 + \frac{n^2}{\zeta}+
                \frac{\xi^2}{\zeta}\right)\right).
          \tag 3-5
$$
This formula is essentially obtained by Winthrop and Worthington
 (\WWT). In the formula, we have also assumed that contribution
from large $n$ is less than that from small $n$ and have
applied the paraxial approximation to large $n$ case.
(This application can be justified by the considerations
that for the case that either $\zeta$ or $\xi$ is
an irrational number, the phase at large $n$
becomes random phase for each $n$ and sum of each terms
with the random phase
damps its amplitude and for the case that both $\zeta$ and $\xi$
are rational numbers,
the net effect from (3-1), in which the larger $n$ contributes less,
repeats to appear in (3-5) but rescaling $C'$ like (A-1) in the
Appendix,
(3-5) brings us the data of the net effect.)

We also consider the same situation as (2-8) or
$\zeta = p/q$ for coprime positive numbers $p$ and $q$.
By introducing the integers $l$, $s$ such that $n=lp+s$,
the quantity of each terms in (3-5) becomes,
$$
        (\xi n + \frac{1}{2}n^2)\frac{q}{p}
     = \xi s\frac{q}{p} + \frac{1}{2}\frac{q}{p}s^2
              + \xi l q + \frac{1}{2} l^2 q p + lq s.
           \tag 3-6
$$
Noting (2-10), (3-5) is expressed by
$$
        \tilde\psi_p(\xi,\zeta)=C'
 \sum_{s = 0}^{q-1}
                 \ee^{\pi i (2 q \xi s + q s^2+ q\xi^2)/p }
\sum_{l = -\infty}^\infty
                 \ee^{\pi i (2 \xi q  + p q) l }.
        \tag 3-7
$$
Corresponding to (2-12), we obtain the wave function of the
system,
$$
        \tilde\psi_p(\xi,\zeta)=C' \sqrt{p}
           \sum_{n= -\infty}^\infty
           \tilde A(n; q, p ) \delta (\xi -\frac{1}{2}e_{qp }
           - \frac{n}{q}  ),
             \tag 3-8
$$
where
$$
         \tilde A(n; q, p ):=\frac{1}{\sqrt{p}}
              \sum_{s = 0}^{q-1}
   \exp\left({i\pi  \left[\left(2 n +
           q e_{qp}\right) s  + q s^2\right]/p +
          \left(2 n +q e_{qp}\right)^2/4pq }\right).
            \tag 3-9
$$
As (2-15), we have
$$
\tilde A(n;q,p) =
\left\{ \matrix
\pmatrix p\\ q\endpmatrix
\exp\left(i
\pi \left[ \dfrac{1}{4}q
        -\left(\dfrac{q}{p}
             \left(\left[\dfrac{1}{q}\right]_{p}\right)^2
               - \dfrac{1}{q p}\right )n^2
  \right]\right), &
             \text{  $p$ even, $q$ odd, } \\
\pmatrix q\\ p\endpmatrix\exp\left( -i
\pi \left[\dfrac{1}{4}(p-1)
+\left(\dfrac{q}{p}
             \left(\left[\dfrac{1}{q}\right]_{p}\right)^2
               - \dfrac{1}{q p}\right )n^2
               \right] \right), &
             \text{  $p$ odd, $q$ even, } \\
\pmatrix q\\ p\endpmatrix \exp
      \left(-i \pi
          \left[ \dfrac{1}{4}(p-1)
             +\left(\dfrac{2q}{p}
                  \left[\dfrac{1}{2} \right]_{p}
                \left[\dfrac{1}{2q}\right]_{p}
             -\dfrac{1}{4q p}\right )(2n+q)^2
          \right]
          \right), &
             \text{  $p$ odd, $q$ odd.}\\
\endmatrix
\right. \tag 3-10
$$
We note that the function form in (3-10) differs from that in
(2-15). Especially
the apparent form of the final case of odd $pq$
is completely different from that in (2-15);
{\it the roles of $q$ and $p$ in both formulae look inverted}.

\vskip 1 cm
{\centerline{\bf{\S 4. Discussion}}}
\vskip 0.5 cm

As we have two expressions of the same Talbot effect,
we will discuss both expressions.

Due to Hecke (\He\ Chap.8), we have the reciprocity of Gauss sums,
if $ab$ even and $c$ is even (or if $ab$ is odd and $c$ is odd)
(\HB),
$$
\sum_{t=0}^{a-1}
\ee^{ \frac{\pi i}{a}(b t^2 + c t ) }
=\left[\frac{ a  i }{ b } \right]^{1/2}
\ee^{ - i \frac{ \pi c^2 }{4 a b}}
\sum_{t=0}^{b-1}
\ee^{ -\frac{\pi i}{b}(a t^2 + c t ) } . \tag 4-1
$$
Here we use the convention $\sqrt{i}=i^{1/2}\equiv \ee^{i\pi/4}$.
From the function forms in (2-15) and (3-10), we have the
relation,
$$
 A(n; q, p )=\sqrt{i} \tilde A(n; q, p )   .\tag 4-2
$$
In other words, the result which we used
the Helmholtz equation
and that used by the Fresnel integration
agree with after applying the paraxial approximation.
Whereas we used the Fourier transformation
on solving the Helmholtz equation,
the computation of Fresnel integration
is directly connected with the optical
paths. As (4-1) is obtained by the
Fourier transformation as mentioned in
the Appendix (A-15) following Appendix in the paper
of Hannay and Berry (\HBT), the correspondence is
very natural. However as in both approaches,
we used the essentially same but different
paraxial approximations, (4-2) is not
trivial. Further by considering the meanings
of (4-2) as follows, it will turn out non-trivial.

Hereafter let us consider the relation (4-2) primitively
in order to reveal the arithmetic structure of (4-2).
First we consider $n$-dependence up to constant factor.
Since $p$ and $q$ are coprime,
$$
\left[\dfrac{1}{q}\right]_{p} q + \left[\dfrac{1}{p}\right]_{q}p
=1 + pq, \tag 4-3
$$
and thus
$$
\left(\left[\dfrac{1}{q}\right]_{p} q \right)^2
+ \left(\left[\dfrac{1}{p}\right]_{q}p\right)^2
\equiv1 +(pq)^2 \quad \text{mod } 2pq. \tag 4-4
$$
Hence the correspondence of
the $n$-dependence parts of even $pq$ case
is expressed by
$$
-\left(\dfrac{q}{p}
  \left(\left[\dfrac{1}{q}\right]_{p}\right)^2
           - \dfrac{1}{q p}\right )n^2
\equiv\dfrac{p}{q}
  \left(\left[\dfrac{1}{p}\right]_{q}\right)^2
           n^2 \quad \text{mod } 2.
  \tag 4-5
$$
Similarly, for the case of $pq$ odd,
we obtain,
$$
\split
-   &\left(\dfrac{2q}{p}\left[\dfrac{1}{2}\right]_{p}
  \left(\left[\dfrac{1}{2q}\right]_{p}\right)^2
        - \dfrac{1}{4q p}\right )(2n+q)^2\\
&\quad \qquad
\equiv\dfrac{2p}{q}\left[\dfrac{1}{2}\right]_{q}
  \left(\left[\dfrac{1}{2p}\right]_{q}\right)^2
          (2n+q)^2 +
 \left ( \frac{1}{2}(p+q+1) +\frac{pq}{4}
\right)(2n+q)^2 \quad \text{mod } 2.
\endsplit
  \tag 4-6
$$
The derivation of (4-6) needs heavy computations in the
sense of primitive number theory due to the factor $1/4$.
We used the relations
for $pq$ odd case,
$$
        \left[ \frac{1}{2}\right]_p = \frac{1+p}{2},
$$
$$
\frac{1}{4}\left(
\left[\dfrac{1}{q}\right]_{p} p + \left[\dfrac{1}{p}\right]_{q}q
\right)
(1+pq) = \frac{p+q}{2} \quad \text{mod }2,
\tag 4-7
$$
which is proved by
by expressing $p=4 \mu + \alpha$ and $q=4\nu +\beta$
($\mu$, $\nu$ are positive integers and
$\alpha$ and $\beta$ are 1, or 3.)
and by considering the possible cases.

With respect to the $n$-dependence,
 except constant phase and factor, we have
the same dependence. It implies that they express
 the same intensity
of light, square of wave function, in the screen.

Next we will consider the equality of (4-2) including the
constant phase factor. The even $pq$ case is not
difficult because (4-5) is essential for the equality.
However for the odd $pq$ case, we need careful
treatment.

Thus we will concentrate our attention on
the odd $pq$ case.
Let us consider the quantity $\tilde A(n,p,q)/A(n,p,q)$
and show that it is $\sqrt{i}$.

Further as for odd $pq$ case, we have
$$
\exp\left(\pi i\left( \frac{1}{2}(p+q+1)
+ \frac{pq}{4}\right)(2n+q)^2\right)
         =
\exp\left(\pi i \left( \frac{1}{2} (p+q+1)+\frac{pq}{4}
\right)\right),
\tag 4-8
$$
by considering possible cases.
Noting (4-6) and (4-8),
the terms $\exp\left( \pi i \left(\dfrac{pq}{4} +\dfrac{p+q+1}{2}
\right)\right)$ multiplying with
 $\exp\left( \pi i\left(  -\dfrac{p-1}{4}\right)\right)$
and $ \exp\left( \pi i \left( -\dfrac{q-1}{4}\right)\right)$
in (2-15) and (3-10) becomes
$$
\exp\left(\pi i \left(\dfrac{(p-1)}{2} \cdot
                        \dfrac{(q-1)}{2} - \frac{1}{4}\right)\right).
\tag 4-9
$$
As it is well-known, the Jacobi's quadratic reciprocity
is given by (\IR, Chap.5)
$$
  \pmatrix p \\ q \endpmatrix \pmatrix q \\ p \endpmatrix
= \exp\left( i\pi\left( \dfrac{(p-1)}{2} \cdot
                        \dfrac{(q-1)}{2}  \right) \right),
 \tag 4-10
$$
It means that (4-9) expresses the Jacobi's
quadratic reciprocity. Hence
 $\tilde A(n,p,q)/A(n,p,q)$
 is $\sqrt{i}$.
 In other words agreement in (4-2) for the
odd $pq$ case is primitively proved.

We note that the agreement in (4-2) essentially comes from
(4-3) and (4-10) and we emphasize that
our correspondence including the sign of the
phase of Gauss sums.
It is known that
determination of the
 sign of the phase in Gauss sums is a very subtle
problem (\IR, p.73).

In truth, we have a natural correspondence including  $\pi/2$
phase in (4-2). The phase $\pi/2$ reminds us of the phase anomaly
(\BW\ 8.8.4)
and caustic problem (\BU).
 (In the paper of Berry and Ustill, we can find a beauty of
another connection between classical optics and modern mathematics.)
The $\pi/2$ phase shift was
studied in context of the partial differential equations
as Maslov connection (\Mas),
 number theory as metaplectic representations (\W),
phase anomaly in optics, connection formula in semi-classical
method in quantum mechanics,
and so on. The {\lq\lq}phase" of the Gauss sums is
a typical example of metaplectic system.
In these systems, the phase is a delicate object.
Since the phase usually appears in optics as a
 phase anomaly, it is expected that
our phase also  appears due to the effects.
Further it is known that in path integral approach,
(or in this case, the Fresnel integral),
the treatment of the phase needs more carefulness
(\Sch\ Chap. 17; \BU; \MatO).
Hence the agreement between both approaches from Helmholtz
equation and Fresnel integral
is highly non-trivial.

However conversely, as we solve the same system by
means of different methods, these physical investigations require
the complete agreement between $\tilde A(n,p,q)$ and $A(n,p,q)$.
It means that {\it our computation gives a novel
 proof of the reciprocity of the Gauss sums
from physical point of view.}

Even though we deal only with the classical optics,
we know a following
correspondence between classical optics and quantum
mechanics. The wave length $\lambda$ is translated into
Planck constant $h$,
the Helmholtz equation corresponds to the Schr\"odinger
equation and the Fresnel integral is related to path integral
(\GS\ I.12; \SW). In paraxial theory of geometrical optics, which
is sometimes called
Gaussian optics (\BW,  4.4), we can find the symplectic structure in
the angle and position of a ray (\GS \ Chap.I; \SW). We can define
a Poisson bracket for the angle and the position
as in the classical mechanics.
In the sense, the wave optics is related to the quantum mechanics.
In truth Hannay and Berry (\HBT) studied the Gauss sums
in the quantum mechanical context, which play the same roles in
Talbot effects in classical optics as mentioned above.

As mentioned in the introduction,
we should regard the result from Helmholtz equation as a wavy
property and one from the Fresnel integral as a particle property.
The complementary principle leads us
that complementary elements, such as configuration and momentum,
are of dual and complementary to each other due to Planck
constant $\hbar$ (\BB).
Though the concept was introduced in order to
explain quantum mechanical experiment (\BB),
this concept might be applicable to
an even purely theoretical phenomena {\it e.g.}, one in number theory.
In fact, in the wave expression (2-7), we find
the term $-2\pi i\zeta \left( \dfrac{1}{2}n^2\right)$
which is proportional to $\lambda$ because $\zeta = z\lambda/a^2$,
whereas in the particle expression (3-5),
$\pi i\left(\dfrac{2\xi n}{\zeta}
+ d\dfrac{n^2}{\zeta}+\dfrac{\xi^2}{\zeta}\right)$
is proportional to $1/\lambda$. When $\lambda$ vanishes,
former one vanishes while the later case diverges.
(In the later case,
the operation is related to another connection between
number theory and quantum system (\MatTT).)
They show the duality and  complementarity.
It implies that the agreement in (4-2) can be
interpreted as  wave-particle complementarity.
Thus {\it the complementarity in this system and
reciprocity of Gauss sums should be regarded
as double aspects of a thing}.

The complementarity between wave and particle,
for the quantum mechanics, is based upon the
commutation relation,
$$
        x p - p x = \sqrt{-1} \hbar, \tag 4-11
$$
for position and momentum operators, $x$ and $p$.
(4-11) is resembles to (4-3), {\it i.e.},
for coprime integers $p$ and $q$, there exists integers
$\left\{\dfrac{1}{q}\right\}_{p}$ and
$\left\{\dfrac{-1}{p}\right\}_{q}$ such that
$$
\left\{\dfrac{1}{q}\right\}_{p} q -
\left\{\dfrac{-1}{p}\right\}_{q}p
=1 . \tag 4-12
$$
(4-12) is very essential and is the most important
origin which generates a beauty of number theory.

Hence this analogy between (4-11) and (4-12)
leads to our conclusion that
the wave-particle complementarity
(in this system) plays the same role as
the reciprocity of coprime numbers in  number theory.

In the studies of mathematical physics, we sometimes
encounter the cases that we feel the resemblances
between number theory and physics as mentioned in the papers
(\MatTT).
The works of Hannay and Berry (\HBT) and Berry and Klein (\BKT)
are the most typical cases.
Further in the optical design, Gaussian optics in lens system can
be expressed by a generalized bracket of
the Gaussian bracket which represents continuous fraction
(\Tan).
In one-dimensional classical point-particle-system,
the cyclomonic polynomial appears and expresses a sort of
 integrable condition (\IMO).
Further in quantum mechanical problems, we can find
the several resemblances. We hope that this report
might have some effects on these studies in future.

\vskip 0.5 cm

{\centerline{\bf{ Appendix}}}

\vskip 0.5 cm

In this appendix, we will base upon the Appendix in the paper of
Hannay and Berry (\HBT)
and supply their arguments. For (2-15) and (3-10), we will show
the derivations of (A-14) and (A-18), which slightly
differ from those in the paper.
The difference assures the equality
in (4-2).

First we will define an infinite sum,
$$
G(a,b,c) := \lim_{N\to \infty}
   \frac{1}{2a b N}
\sum_{m=-Nab}^{Nab}
\exp\left(
 \dfrac{i\pi  }{b}\left[ am^2 + cm \right] \right),
\tag A-1
$$
and the Gauss sum,
$$
K(a,b,c) :=
   \frac{1}{a}
\sum_{m=0}^{a-1}
\exp\left(
 \dfrac{i\pi  }{b}\left[ am^2 + cm \right] \right).
\tag A-2
$$
By letting $m = b n + s$, each term in (A-1) becomes
$$
\exp\left(
\frac{\pi i}{b} (a m^2 + c m ) \right) =
\exp\left(
\frac{\pi i}{b} (a s^2 + c s )\right),
\tag A-3
$$
if $ab$ and $c$ are even (or $ab$ and $c$ are odd).
Hence if $ab$ and $c$ are even (or $ab$ and $c$ are odd),
(A-1) and (A-2) coincide with,
$$
G(a,b,c) \equiv K(a,b,c). \tag A-4
$$

For the case of that $a$ and $c$ are even and $a$ and $b$ are
coprime, we can compute $K(a,b,c)$ as,
$$
\split
K(a,b,c) &=   \frac{1}{b}
\sum_{m=0}^{b-1}
\exp\left(
 \dfrac{2\pi i a/2}{b}\left[ m+\dfrac{c}{2}
        \left[\dfrac{1}{a}\right]_b \right]^2 \right)
\exp\left(
 \dfrac{-\pi i a}{b}\left[\dfrac{c}{2}\right]^2
\left[\dfrac{1}{a}\right]_b^2 \right)\\
&= \frac{1}{b}
\sum_{n=0}^{b-1}
\left[1 + \pmatrix a n/2 \\ b \endpmatrix\right]
\exp\left(
 \dfrac{2\pi i }{b} n \right)
\exp\left(
 \dfrac{-\pi i a}{b}\left[\dfrac{c}{2}\right]^2
\left[\dfrac{1}{a}\right]_b^2 \right)\\
& =\frac{1}{b}\pmatrix a /2 \\ b \endpmatrix
\exp\left(
 \dfrac{-\pi i (a/2)}{b}\left[\dfrac{c}{2}\right]^2
\left[\dfrac{1}{a}\right]_b^2 \right)
\sum_{n=0}^{b-1}
\pmatrix  n \\ b \endpmatrix
\exp\left(
 \dfrac{2\pi i }{b} n \right)
.
\endsplit\tag A-5
$$
In the case for odd $b$ (\IR, Chap.6), the Gauss sum is expressed by,
$$
\split
\sum_{n=0}^{a-1}
\pmatrix  n \\ b \endpmatrix
\exp\left( \dfrac{2\pi i }{b} n \right)
& = \left\{
    \matrix
       \sqrt{b},  & \text{for } b =  1 \text{ mod } 4,\\
       i\sqrt{b},  & \text{for } b =  3 \text{ mod } 4.
     \endmatrix
\right. \\
&= \sqrt{b} \exp\left( \frac{i \pi}{8}(b-1)^2 \right).
\endsplit
\tag A-6
$$
Using the relations,
$$
        \pmatrix 2 \\ b\endpmatrix =
                \pmatrix 1/ 2 \\ b\endpmatrix = (-1)^{(b^2-1)/8},
\quad
        \pmatrix -1 \\ b\endpmatrix  = (-1)^{(b-1)/2},
            \tag A-7
$$
we have
$$
        \pmatrix 2 \\ b\endpmatrix
\sum_{n=0}^{a-1}
\pmatrix  n \\ b \endpmatrix
\exp\left(
 \dfrac{2\pi i }{b} n \right) = \sqrt{b}
\exp\left(\frac{-i \pi}{4}(b-1) \right),
$$
$$
        \pmatrix -2 \\ b\endpmatrix
\sum_{n=0}^{a-1}
\pmatrix  n \\ b \endpmatrix
\exp\left(
 \dfrac{2\pi i }{b} n \right) = \sqrt{b}
\exp\left(\frac{i \pi}{4}(b-1) \right).
 \tag A-8
$$

Next we will consider the case
that $ab$ and $c$ are odd and $a$ and $b$ are
coprime.
If we deal with $\exp( 2\pi i  m/b )$-type
functions, we can use a technique mentioned
in the paper of Hannay and Berry (\HBT), {\it i.e.},
$\sum_{m=1}^{b-1} \exp( 2\pi i a m/b )=$
$\sum_{m=1}^{b-1} \exp( 2\pi i  m/b )$ for
coprime and odd $ab$.
However we are treating $\exp( \pi i  m/b )$-type
functions.
Since the period of $\exp( \pi i  m/b )$ in $m$
is not $b$ but $2b$, we must independently
consider this case, which is completely different
from the even $a$ case.

Since $b+1$ is even and $b+1\equiv 1$ modulo $b$, we have
$$
 \left[\frac{1}{2}\right]_b = \frac{ 1 + b}{2}.
\tag A-9
$$
Hence we have
$$
        \exp\left( \pi i \frac{a}{b} m^2\right)
           = \exp\left( 2 \pi i  \left[\frac{1}{2}\right]_b
            \frac{a}{b}m^2 - \pi
 i a m^2 \right), \tag A-10
$$
and
$$
        \exp\left( \pi i \frac{c}{b} m\right)
           = \exp\left( 2 \pi i  \left[\frac{1}{2}\right]_b
            \frac{c}{b}m - \pi
 i c m \right). \tag A-11
$$
Noting $(-1)^{am^2}\equiv (-1)^{am}$,
$$
        \exp\left( -\pi i a  m^2 + \pi i c m \right)=1.\tag A-12
$$
Accordingly for odd $ab$ and odd $c$ case, we have,
$$
\split
K(a,b,c) &=   \frac{1}{b}
\sum_{m=0}^{b-1}
\exp\left(
 \dfrac{2\pi i a}{b}\left[\frac{1}{2}\right]_b\left[ m+c
        \left[\dfrac{1}{2a}\right]_b \right]^2 \right)
\exp\left(
 \dfrac{-2\pi i a}{b} c^{2}\left[\dfrac{1}{2}\right]_b
\left[\dfrac{1}{2a}\right]_b^2 \right)\\
& =\frac{1}{b}\pmatrix a/2 \\ b \endpmatrix
\exp\left(
 \dfrac{-2\pi i a}{b} c^{ 2}
\left[\dfrac{1}{2}\right]_b
\left[\dfrac{1}{2a}\right]_b^2 \right)
\sum_{n=0}^{b-1}
\pmatrix  n \\ b \endpmatrix
\exp\left(
 \dfrac{2\pi i }{b} n \right)\\
& =\frac{1}{\sqrt{b}}\pmatrix a  \\ b \endpmatrix
\exp\left(\dfrac{-\pi i }{4b}(b-1)
          -\dfrac{2\pi i a }{b}c^{ 2}
          \left[\dfrac{1}{2}\right]_b
         \left[\dfrac{1}{2a}\right]_b^2
    \right).
\endsplit\tag A-13
$$

Here following the argument in the paper of Hannay and Berry (\HBT),
we will prove the reciprocity of the
infinite sums for any $a$, $b$ and $c$,
$$
G(a,b,c) = G(-b,a,c) \sqrt{\dfrac{ia}{b}}
\exp\left[\frac{-i \pi c^2}{4ab}\right]. \tag A-14
$$
We first note the relations;
$$
\frac{1}{\sqrt{2\pi}}\int d x \ee^{-\beta x^2} \ee^{i kx}
= \sqrt{\frac{1}{\beta}}\ee^{-k^2 / 4\beta}. \tag A-15
$$
$$
\split
        \delta_c(x) &= \sum_{m=-\infty}^\infty \delta(x-m)\\
         & =\frac{1}{2\pi}\sum_{m=-\infty}^\infty
             \int_{-\infty}^{\infty} d k \ee^{ik (x-m)}\\
           &= \sum_{m=-\infty}^\infty  \ee^{2\pi i xm}.
\endsplit
         \tag A-16
$$
Secondly, we note that due to the paper of Hannay and Berry (\HBT),
the left hand side in (A-14) is expressed by,
$$
G(a,b,c) = \lim_{\alpha\to +0}
 \int_{-\infty}^\infty \sqrt{\alpha}
\ee^{-\pi \alpha x^2}
\delta_c(x) \exp\left(
\frac{i\pi}{b}(ax^2+cx)\right) dx. \tag A-17
$$
Using (A-15) and (A-16), we compute (A-17) and then derive
(A-14). Using (A-4), (A-14) becomes the  reciprocity of the
Gauss sums.

Hence the Gauss sums are expressed by,
$$
K(a,b,c)  =
\left\{ \matrix
\dfrac{1}{\sqrt{b}} \pmatrix a\\ b\endpmatrix
\exp\left(-i
\pi \left[ \dfrac{1}{4}(b-1)
         + \dfrac{a}{b}
             \left(\left[\dfrac{1}{a}\right]_{b}\right)^2
            \left(\dfrac{c}{2}\right)^2
  \right]\right), &
             \text{  $a$ even, $b$ odd, } \\
\dfrac{1}{\sqrt{b}} \pmatrix b\\ a\endpmatrix
\exp\left(i
\pi \left[ \dfrac{1}{4}a
         - \dfrac{a}{b}
             \left(\left[\dfrac{1}{a}\right]_{b}\right)^2
                \left(\dfrac{c}{2}\right)^2
               \right] \right), &
             \text{  $b$ odd, $a$ even, } \\
\dfrac{1}{\sqrt{b}} \pmatrix a\\ b\endpmatrix
\exp\left(-i
\pi \left[ \dfrac{1}{4}(b-1)
        + \dfrac{2a}{b}
             \left[\dfrac{1}{2}\right]_{b}
           \left(\left[\dfrac{1}{2a}\right]_{b}\right)^2
                c^2
             \right]\right), &
             \text{  $a$ odd, $b$ odd.}\\
\endmatrix
\right.\tag A-18
$$

{\centerline{\bf{Short titile for page headings:}}}

{\centerline{\bf{Complementarity and Reciprocity on Talbot Effects}}}

\enddocument